\documentclass[prd,preprint,
superscriptaddress,
nofootinbib,%
tightenlines
]{revtex4}
\usepackage{epsfig}
\usepackage{color}
\usepackage{slashed}
\usepackage{amssymb}

\newcommand{\ben}{\begin{displaymath}}
\newcommand{\een}{\end{displaymath}}
\newcommand{\be}{\begin{equation}}
\newcommand{\ee}{\end{equation}}
\newcommand{\bea}{\begin{eqnarray}}
\newcommand{\eea}{\end{eqnarray}}
\begin{document}

\title{Meson-baryon scattering in resummed baryon chiral perturbation theory using time-ordered perturbation theory}
\author{X.-L.~Ren}
 \affiliation{Ruhr University Bochum, Faculty of Physics and Astronomy,
Institute for Theoretical Physics II, D-44870 Bochum, Germany}
\author{E.~Epelbaum}
 \affiliation{Ruhr University Bochum, Faculty of Physics and Astronomy,
Institute for Theoretical Physics II, D-44870 Bochum, Germany}
\author{J.~Gegelia}
\affiliation{Ruhr University Bochum, Faculty of Physics and Astronomy,
Institute for Theoretical Physics II, D-44870 Bochum, Germany}
\affiliation{Tbilisi State  University,  0186 Tbilisi,
 Georgia}
\author{Ulf-G.~Mei{\ss}ner}
 \affiliation{Helmholtz Institut f\"ur Strahlen- und Kernphysik and Bethe
   Center for Theoretical Physics, Universit\"at Bonn, D-53115 Bonn, Germany}
 \affiliation{Institute for Advanced Simulation, Institut f\"ur Kernphysik
   and J\"ulich Center for Hadron Physics, Forschungszentrum J\"ulich, D-52425 J\"ulich,
Germany}
\affiliation{Tbilisi State  University,  0186 Tbilisi, Georgia}

\date{16 March, 2020}

\begin{abstract}
Integral equations for meson-baryon scattering amplitudes are obtained by utilizing time-ordered
perturbation theory for a manifestly Lorentz-invariant formulation of baryon chiral perturbation theory. 
Effective potentials are defined as sums of  two-particle irreducible contributions of time-ordered diagrams and
the scattering amplitudes are obtained as solutions of integral equations.
Ultraviolet renormalizability is achieved by solving integral equations for the leading order amplitude 
and including higher order corrections perturbatively.  
As an application of the developed formalism,  pion-nucleon scattering  is considered.

\end{abstract}

\maketitle

\section{\label{introduction}Introduction}

Understanding  meson-baryon scattering processes at low and intermediate energies involving strangeness is a
non-trivial problem. One often uses the so-called chiral unitary approach (see e.g.~Ref.~\cite{Oller:2000ma} for an
early review), which involves a non-perturbative resummation of the chiral amplitude to extend the range of
applicability of low-energy effective field theory (EFT) into the resonance region. This, however, comes with certain shortcomings
as discussed below. So far, a variety of unitarization methods have been proposed. In the pioneering work by
the Munich group~\cite{Kaiser:1995eg,Kaiser:1995cy,Kaiser:1996js}, the Lippmann-Schwinger equation in coupled
channels were used to iterate the leading order (LO) kernel in the region of the $\Lambda(1405)$ and $N^*(1535)$
resonances, employing Gaussian regulators to tame the UV behaviour of the chiral potential.
In Refs.~\cite{Oset:1997it,Oller:1997ti}, the relativistic counterpart of the scattering equation, the
Bethe-Salpeter equation, was employed to sum up the iterations of the covariant interaction kernel.
Besides, different  frameworks were subsequently developed based on the inverse amplitude
method (IAM)~\cite{Oller:1998hw,GomezNicola:1999pu,GomezNicola:2000wk},
the N/D method (based on dispersion relations) \cite{Meissner:1999vr,Oller:2000fj} and the Bethe-Salpeter
equation supplemented by large-$N_C$ constraints~\cite{Lutz:2001yb}, to name a few. Clearly, having such a variety
of unitarization schemes introduces some model-dependence in the  chiral unitary approach,
which can be minimized if one enforces a matching to the perturbative amplitudes, as suggested in~\cite{Oller:2000fj}.

Chiral unitary approaches have been used to describe hadron scattering amplitudes and interpret
the molecular components of resonances. Arguably the most striking result of this method is the
interpretation of the $\Lambda(1405)$ resonance as a superposition of two states \cite{Oller:2000fj,Jido:2003cb}.  
To identify the nature of resonances, one has to carefully derive the kernel of the meson-baryon scattering amplitude
order by order in chiral perturbation theory (ChPT). At lowest order,
one has to take into account the Weinberg-Tomozawa (WT) contact
term as well as the Born and crossed-Born term contributions. Often considered as the most important piece of the LO kernel,
the WT term has been mostly employed in the initial studies of meson-baryon scattering, see e.g.~Ref.~\cite{Oset:1997it}. 
This approximation should, however, not be performed any more. First,  one cannot expect that this is suitable
for all the meson-baryon scattering channels, because the WT term does not contribute to  e.g.~the
$K^- p \to K^+\Xi^-, K^0\Xi^0$ reactions. Second, and most importantly, such
an approach violates the counting rules of the
underlying effective field theory, which states that one has to include all terms at a given order,
not just picking the presumably dominant one(s). Thus, beyond the WT term, the Born and crossed-Born terms
in the lowest order and the higher order contributions are necessary to improve the description of the rich
information available for meson-baryon scattering.

Along this line, in Ref.~\cite{GomezNicola:1999pu}, the scattering amplitudes up to next-to-next-to-leading order
(NNLO) in the
heavy baryon (HB) ChPT~\cite{Jenkins:1990jv,Bernard:1992qa} have been
employed in the chiral unitary approach. The obtained results turned
out to provide a reasonably good description of the scattering data up to around $1.3$~GeV,
including the region of the $\Delta(1232)$ resonance in the $P_{33}$ partial wave. A further step in this direction
was the
Bethe-Salpeter approach in Ref.~\cite{Bruns:2010sv} (see also Ref.~\cite{Nieves:1999bx})
used to investigate pion-nucleon scattering in the $S_{11}$ partial wave, showing that both the $N^*(1535)$ and
the $N^*(1650)$ can be dynamically generated. Also, it should be pointed out that state-of-the-art investigations of the
$\Lambda(1405)$ employ kernels at least to NLO accuracy, see e.g.~Ref.~\cite{Cieply:2016jby}
for a comparison of different approaches and Ref.~\cite{Sadasivan:2018jig} for a recent study including also $P$-waves.
Extending these results beyond NLO accuracy
can  indeed lead to distortions of the analytic structure, as exemplified in Ref.~\cite{Yao:2015qia}.

In recent years, covariant ChPT with the extended-on-mass-shell (EOMS)
scheme~\cite{Gegelia:1999gf,Gegelia:1999qt,Fuchs:2003qc} was utilized because of the somewhat faster convergence than
HB scheme in the one-baryon sector \cite{Alarcon:2012kn,Chen:2012nx,Yao:2016vbz,Lu:2018zof}. Hence,
in the chiral unitary approach, one might also want to use the relativistic meson-baryon interaction from
covariant ChPT.  Thus, the relativistic integral equation, e.g.~the Bethe-Salpeter equation ($T=V+ VGT$), has to
be employed to obtain the unitarized amplitude in such a  Lorentz-invariant framework. This is, in general,
a technically very demanding  task. In practice, the approximation of  on-shell factorization,
which takes $V$ and $T$ on shell to factor out the four-dimensional integral, is often used to solve the
Bethe-Salpeter equation~\cite{Oset:1997it,Oller:1998zr}.

Due to the resummation of the interaction kernel in the unitarization procedure, not all the ultraviolet
divergent terms of the meson-baryon scattering amplitude can be absorbed in the low-energy constants (LECs)
of the effective Lagrangians.  Therefore, in chiral unitary approaches, the amplitudes depend on the cutoff
parameter~($\Lambda$) or the subtraction constant(s)~\cite{Oller:2000fj,Borasoy:2007ku,Hyodo:2011ur}.
To obtain an explicitly renormalizable approach we apply the rules of time-ordered perturbation theory (TOPT)
\cite{Baru:2019ndr} to the effective Lagrangian of mesons, baryons and vector mesons as dynamical degrees
of freedom. The inclusion of vector mesons leads to a softer
UV-behaviour as will be discussed below.
We define the effective meson-baryon potential as the sum of the two-particle irreducible TOPT diagrams
contributing to the meson-baryon scattering amplitudes. 
The scattering amplitudes are obtained by solving the corresponding integral equations.
The advantage of this formulation as compared to the alternative
approaches mentioned above is that the leading-order scattering
amplitude is renormalizable. 
This guarantees that all divergences can be removed by renormalizing the coupling
constants available at a given order, provided that the higher-order corrections to the effective potential are
taken into account perturbatively. To demonstrate how this formalism can be applied to a meson-baryon
scattering problem, we apply it to elastic pion-nucleon ($\pi N$)  scattering, where we use the
parameterization of fields specified in Ref.~\cite{Weinberg:1968de} (this parametrization is only suitable
for the two-flavor case).

Our paper is organized as follows: In section~\ref{effective_Lagrangian} we
specify the effective Lagrangian for meson-baryon scattering in the three-flavor case.
An integral equation for the meson-baryon scattering amplitude using TOPT is derived 
in section~\ref{intequation}.  In section~\ref{pN}, we discuss the application of the developed formalism to
the LO pion-nucleon scattering amplitude and the results of our work are
summarized in  section~\ref{conclusions}.  Some technicalities are relegated to the appendices.

\section{Effective Lagrangian}
\label{effective_Lagrangian}

We start with the effective Lagrangian of the interacting SU(3) octet fields  of pseudoscalar mesons $P$, baryons $B$, 
and the vector mesons $V_\mu$ in the vector field representation of Ref.~\cite{Borasoy:1995ds} (corresonding to 
model II of Ref.~\cite{Ecker:1989yg}) invariant under the symmetries of QCD, in particular the non-linearly realized
spontaneously broken chiral symmetry. We include vector mesons as explicit degrees of freedom because this
improves the ultraviolet behaviour of meson-baryon integral equations without altering the low-energy scattering
amplitudes. However, care has to be taken to avoid double counting, exemplified for the WT term in different
effective Lagrangians in Ref.~\cite{Borasoy:1995ds}.


Our lowest-order Lagrangian is given by
\begin{eqnarray}
{\cal L}_0 &=& \frac{F_0^2}{4}\, \mbox{Tr}\left\{ u_\mu
u^\mu +\chi_+\right\} 
+ \mbox{Tr} \left\{ \bar {\rm B} \left( i\gamma_\mu D^\mu -m \right)  {\rm B} \right\}  
- \frac{1}{4} \,  {\rm Tr} \left( V_{\mu\nu}  V^{\mu\nu} - 2 M_V^2 \,V_{\mu} V^{\mu} \right)  \nonumber\\
&+& \frac{D/F}{2}
\mbox{Tr} \left\{\bar {\rm B} \gamma_\mu \gamma_5 [u^\mu,{\rm B}]_{\pm}\right\} +\left( G_D/G_F\right)
\mbox{Tr} \left\{\bar {\rm B} \gamma_\mu [V^\mu,{\rm B}]_{\pm}\right\}, 
\label{LagrSU3}
\end{eqnarray}
where 
\begin{eqnarray}
&& V_{\mu\nu}= D_\mu V_\nu - D_\nu V_\mu, \ D_\mu{\rm X} = \partial_\mu {\rm X}+[\Gamma_\mu , {\rm X}], \ \Gamma_\mu =\frac{1}{2} \left( u^{\dagger} \partial_\mu u + u  \partial_\mu u^\dagger \right), \nonumber\\
&& u_\mu =iu^{\dagger} \partial_\mu U u^{\dagger},\ u^2=U=\exp\left(\sqrt{2} i P/ F_0 \right) ,\ 
\chi_{\pm}= u^\dagger\chi u^\dagger
\pm u \chi^\dagger u, \  \chi = 2 B_0 {\cal M}~.  
\end{eqnarray}
Here, $F_0$ is the pion decay constant in the three-flavor chiral limit, while $D$, $F$, $G_D$ and $G_F$
are coupling constants, ${\cal M}$ denotes the quark mass matrix and
$B_0$ is related to the scalar quark condensate. The SU(3) matrix $U$
is parametrized in terms of the pseudoscalar meson octet.  

We take into account the results of Ref.~\cite{Unal:2015hea} obtained from the analysis of constraints imposed 
on the interactions of vector meson fields leading to $G_D=0$ and $G_F=g$, with  $g$ the coupling of the
vector-field self-interactions, corresponding to a massive Yang-Mills theory~\cite{Meissner:1987ge,Birse:1996hd}.
Analogously to Ref.~\cite{Ecker:1989yg}, we introduce new vector fields by substituting $V_\mu
= \bar V_\mu -(i/g) \Gamma_\mu$ and obtain, modulo terms of higher order in the chiral expansion and/or
with more than two vector fields, the following Lagrangian
\begin{eqnarray}
{\cal L}_0 &=& \frac{F_0^2}{4}\, \mbox{Tr}\left\{ u_\mu
u^\mu +\chi_+\right\} 
+ \mbox{Tr} \left\{ \bar {\rm B} \left( i\gamma_\mu \partial^\mu -m \right)  {\rm B} \right\}  \nonumber\\
&-& \frac{1}{4} \,  {\rm Tr} \left( \bar V_{\mu\nu}  \bar V^{\mu\nu} - 2 M_V^2 \,\left( \bar V_{\mu} -\frac{i}{g} \, \Gamma_\mu\right)  \left( \bar V^{\mu} -\frac{i}{g} \, \Gamma^\mu\right)  \right)  \nonumber\\
&+& \frac{D/F}{2}
\mbox{Tr} \left\{\bar {\rm B} \gamma_\mu \gamma_5 [u^\mu,{\rm B}]_{\pm}\right\} + g \,
\mbox{Tr} \left\{\bar {\rm B} \gamma_\mu [\bar V^\mu,{\rm B}]\right\} ,
\label{LagrSU3New}
\end{eqnarray}
where $\bar V_{\mu\nu} = \partial_\mu \bar V_\nu - \partial_\nu \bar V_\mu -i g[\bar V_\mu, \bar V_\nu]$.
Notice that the covariant derivatives have been replaced by ordinary ones in Eq.~(\ref{LagrSU3New}),
similar to the two-flavor  parameterization of  Ref.~\cite{Weinberg:1968de}. 
To calculate the meson-baryon scattering amplitudes, we apply the diagrammatic rules  of TOPT~\cite{Baru:2019ndr}
corresponding to the effective Lagrangian of Eq.~(\ref{LagrSU3New}). 

We should mention here that this approach still lacks some physics, namely the explicit inclusion of
the $\Delta(1232)$ resonance, see e.g.~Ref.~\cite{Meissner:1999vr} (or, more generally, the inclusion
of the spin-3/2 decuplet). The $\Delta(1232)$ can not be
generated dynamically if one insists on a matching to chiral amplitudes at low energies or in the
unphysical region as done in
Refs.~\cite{Buettiker:1999ap,Hoferichter:2015tha,Siemens:2016jwj}.
In particular, very large dimension-two and dimension-three LECs
incompatible with the above mentioned determinations were found to be necessary
in order  to generate a resonance in the $P_{33}$-wave using the  IAM
in Ref.~\cite{GomezNicola:1999pu}.

\section{Integral equations for meson-baryon scattering}
\label{intequation}

The meson-baryon scattering amplitude $T_{MB}$ is obtained from the four-point
vertex function $\tilde\Gamma_{4}$ by applying the standard LSZ formula
\begin{equation}
T_{MB} =Z_{B_i}^{1/2} Z_{B_f}^{1/2} Z_{M_i}^{1/2}Z_{M_f}^{1/2}\, \bar u(p_f)\,\tilde\Gamma_{4}^{}\,u(p_i)
\equiv Z_{B_i}^{1/2} Z_{B_f}^{1/2} Z_{M_i}^{1/2}Z_{M_f}^{1/2}\, \tilde T~,
\label{LSZtosh}
\end{equation}
where $Z_{M_i}$ ($Z_{M_f}$) and $Z_{B_i}$ ($Z_{B_f}$) are the residues of the propagators corresponding to the
initial (final) meson and baryon, respectively  and $u$, $\bar u$
are Dirac spinors corresponding to the incoming and outgoing baryons, in order. 
The on-shell amplitude $\tilde T$  is given as a sum of an
infinite number of TOPT diagrams. Notice that it does not include
diagrams with corrections on the external legs.  
Let us discuss this  in more detail. It is convenient to 
define the effective meson-baryon potential as a sum of all possible 
meson-baryon irreducible TOPT diagrams. 
The amplitude $\tilde T$  is then given by an infinite series
\begin{eqnarray}
\tilde T &=& \tilde V+\bar V G\, \bar V +\bar V G\, V G\, \bar V+ \bar V G\, V G\, V G\, \bar V
+\cdots \nonumber\\
& = & \tilde V+\bar V G\, \bar V +\bar V G\, \left[ V + V G\, V  +\cdots \ \right] G\, \bar V
= \tilde V+\bar V G\, \bar V +\bar V G\, T G\, \bar V,
\label{Tseries}
\end{eqnarray}
where $G$ is the meson-baryon Green function and $\tilde T$,
$T$, $\tilde V$, $\bar V$ and $V$  are the on-shell amplitude, 
the off-shell amplitude, the on-shell potential, the half-off-shell
potential and the off-shell potential,   respectively. The on-shell
potential $\tilde V$ does not include diagrams with corrections on the
external legs. The half-off-shell potential $\bar V$ does not include
diagrams with corrections on the external legs with on-shell momenta while
the off-shell potential $V$ also includes diagrams with corrections on
the external legs.   
The off-shell amplitude $T$  satisfies the following equation: 
\begin{equation}
T=V+ V G\,T\,.
\label{Teq1}
\end{equation}
To cover all processes with different strangeness, Eq.~(\ref{Teq1}) has to be understood as a matrix
equation, i.e. one has to deal with coupled channels.

The meson-baryon scattering amplitude can be conveniently calculated in
the center-of-mass system (CMS).  
We denote 
the relative three-momenta of the incoming and
outgoing particles  in the CMS by  $\vec p$ and  $\vec p\,'$, respectively.
In the partial wave basis, Eq.~(\ref{Teq1}) leads to
the following coupled equations with the potentials
$V^{M_fB_f,M_iB_i}\left(\vec p\,', \vec p\right)$,
\begin{eqnarray}
T^{M_fB_f,M_iB_i}\left(E; {\vec p\,'}
,\vec p\right) &=& V^{M_fB_f,M_iB_i}\left(E;\vec p\,',\vec p\right)   \nonumber\\
& +& \sum_{M,B}
\int
\frac{d ^3\vec k }{(2 \pi)^3} V^{M_fB_f,MB}(E;\vec p\,',\vec k)\, G^{MB}(E) \, T^{MB,M_iB_i}
(E; \vec k ,\vec p),
\label{PWEHDR}
\end{eqnarray}
where $M_iB_i,M_fB_f$ and $MB$ denote initial, final and intermediate
particle channels.  Further, the two-body Green functions read
\begin{equation}
G^{MB}(E)= \frac{1}{2\, \omega_M \omega_B}\, \frac{-m_B}{E - \omega_M-\omega_B+i \epsilon} \,,
\label{Gij}
\end{equation}
where $m_I$ and $\omega_I\equiv \omega_I(q,m_I):=\left(\vec q\,{  }^2+m_I^2\right)^{1/2}$ are  the mass and
energy of  the $I^{\rm th}$ hadron.

\medskip

To calculate the meson-baryon scattering amplitudes, we apply the standard power counting to the effective
potential for its expansion in powers of a small parameter and solve the leading order equation for the amplitude
\begin{equation}
T_0 = V_0 + V_0 \, G \ T_0 \,.
\label{EqForGFLO}
\end{equation}
Higher order corrections to the effective potential can be taken into account perturbatively, or alternatively,
subtractive renormalization, analogous to the one outlined in Ref.~\cite{Epelbaum:2020maf}, can be applied.
For the next-to-leading order correction $T_1$ we have
\begin{equation}
T_1=V_1+ T_0\,{ G}\, V_1+ V_1\,{ G}\,
T_0+ T_0\,{ G}\,V_1\,{ G}\,T_0\,,
\label{NLOeqsolution}
\end{equation}
and higher order corrections can be obtained analogously. In practice,  we will solve the half-on-shell equation
and then put  the solution  fully on-shell. 

In the next section we apply this  formalism to  $\pi N$ scattering as an example. Applications in SU(3)
BChPT will be considered in forthcoming publications.

\section{Application to pion-nucleon scattering}
\label{pN}

In the  limit of exact isospin symmetry, the on-shell amplitude of
the elastic $\pi N$ scattering reaction $\pi^a(q_1)+N(p_1)\to\pi^{b}(q_2)+N(p_2)$,
with Cartesian isospin indices $a$ and $b$,
can be parameterized as
\bea
T_{\pi N}^{b a}(s,t,u)=\chi_{N^\prime}^\dagger\left\{\delta_{b a}T^+(s,t,u)
+\frac{1}{2}[\tau_{b},\tau_a]T^-(s,t,u)\right\}\chi_N\ ,
\eea
where the $\tau_i$ are the Pauli matrices and $\chi_{N}$, $\chi_{N^\prime}$ denote nucleon iso-spinors. The 
conventional Mandelstam variables are defined as $s=(p_1+q_1)^2,\ t=(p_1-p_2)^2,\ u=(p_1-q_2)^2$,
subject to the constraint $s+t+u= 2(m_N^2+M_\pi^2)$.

The Lorentz decomposition of the invariant amplitudes $T^{\pm}$ reads (we use here the $D$-$B$
representation instead of the more common $A$-$B$ one, see e.g.~\cite{Hoehler}),
\bea\label{eq.DB}
T^{\pm}(s,t,u)=\bar{u}^{(\lambda^\prime)}(p_2)\left\{D^\pm(s,t,u)
-\frac{1}{4m_N}[\slashed{q}_2,\slashed{q}_1]B^\pm(s,t,u)\right\}u^{(\lambda)}(p_1)\ ,
\eea
with the superscripts $\lambda^\prime$, $\lambda$ denoting the spins of the Dirac spinors
$\bar{u}$, $u$, respectively. 

We use Dirac spinors $u(p)$ with four-momentum $p$:
\be\label{nuclspin}
 u(p) = \left(\frac{\omega(p,m)+m}{2m}\right)^{1/2}    \left(\begin{array}{c} \chi \\
                        \frac{\vec{\sigma}\cdot\vec{p} \ \chi }{\omega(p,m)+m}
                 \end{array}\right), 
\ee 
where $m$ is the mass of the corresponding baryon and $\chi$ a two-component spinor, and decompose 
\begin{eqnarray}
u(p) & = & u_0 +\left[ u(p)-u_0\right] 
\equiv u_0 + u_{\rm ho}\,.
\end{eqnarray}
Here $u_0= \left( \chi \;  0 \right)^T $ is the leading order contribution and  $ u_{\rm ho}$  stands for the higher order part. 
The leading order contribution satisfies 
\begin{eqnarray}
u_0 &=& P_+\, u_0 := \frac{1+v \hspace{-.55em}/\hspace{.1em}}{2} \,u_0\,, 
\label{spellout}
\end{eqnarray}
with $v=(1,0,0,0)$ in the rest-frame of the particle.
For the reduced amplitude we use the following parameterization \cite{Fettes:1998ud}
\bea
T^{ba}_{\pi N}= 
\delta^{ba} \left[ g^+
+ i \,\vec \sigma \cdot(\vec q_2\times \vec q_1) 
 h^+  \right] + i\,\epsilon^{bac} \tau^c \left[ g^-
+ i \,\vec \sigma \cdot(\vec q_2\times \vec q_1) 
 h^-  \right] .
 \label{FMAmp}
\eea

The partial wave projection of the isospin amplitudes is given by
\bea
f_{\ell\pm}^\pm(s)  = \frac{m_N}{8\pi\sqrt{s}}  \int_{-1}^{+1}  dz\left[ g^\pm \,P_\ell(z) +q(s)^2
h^\pm \,\left( P_{\ell\pm 1}(z) -z P_\ell(z)\right)  \right],
\quad z\equiv\cos\theta ,
\eea
where $\theta$ is the scatting angle in the
CMS frame, the
$P_\ell(z)$  are the Legendre polynomials and $q(s)^2=((s-m_N^2-M_\pi^2)^2-4 m_N^2 M_\pi^2)/({4\,s})$. A commonly
used parametrization of the partial wave amplitudes is
 \bea\label{eq:Smatrix}
f_{\ell\pm}^I(s)&=& 
\frac{1}{2iq(s)}\left\{e^{2i\delta_{\ell\pm}^I(s)}-1\right\}.
 \eea
Here, the phase shifts $\delta_{\ell\pm}^I(s)$ are real-valued functions.

\subsection{LO pion-nucleon potential}
We take the effective Lagrangian of pions, nucleons and the $\rho$-meson contributing to the LO $\pi N$
potential in the form given by Weinberg in Ref.~\cite{Weinberg:1968de}, where we also use the universality
of the $\rho$-meson coupling  \cite{Djukanovic:2004mm}:
\begin{eqnarray}
{\cal L} &=& \frac{1}{2}\,\partial_\mu \pi^a \partial^\mu \pi^a -\frac{M^2}{2}\, \pi^a \pi^a  +\bar \Psi \left( i\gamma_\mu \partial^\mu -m + \frac{g}{2} \, \gamma^\mu \tau^a \, \rho_\mu^a
+\frac{1}{2} \stackrel{\circ}{g_A}\gamma_\mu \gamma_5 u^\mu\right) \Psi, 
\nonumber \\
&-& \frac{1}{4} \, F^a_{\mu\nu}  F^{a \mu\nu} + \frac{M_\rho^2}{2}\,\rho_\mu^a \rho^{a\mu} + g \, \epsilon^{abc} \pi^a \partial_\mu \pi^b \rho^{c\mu}\,.
\label{lolagr}
\end{eqnarray}
Here,  $\pi^a$ and $\rho^a_\mu$ are iso-triplets of the pion and $\rho$-meson fields with masses $M$ and
$M_\rho$, respectively, and $M_\rho^2=2 g^2 F_\pi^2$ (KSFR relation)\footnote{Here, we identify the LO
  pion decay constant $F_0$, see Eq.~(\ref{LagrSU3}), with the physical pion decay constant $F_\pi$.}. Further,
$F^a_{\mu\nu} =\partial_\mu \rho_\nu^a-\partial_\nu \rho_\mu^a+ g\, \epsilon^{abc} \rho_\mu^b \rho_\nu^c$,
$\Psi$ is the doublet of the nucleon fields, 
$m$ and $\stackrel{\circ}{g_A}$ are the chiral limit values of the nucleon mass and the axial-vector coupling
constant, respectively.  Notice that by integrating out the vector mesons from EFT defined by the Lagrangian
of  Eq.~(\ref{lolagr}), one generates the standard chiral effective Lagrangian of pions and nucleons alone,
including the  Weinberg-Tomozawa term. As mentioned above, we prefer to work with dynamical vector mesons
because the vector meson exchange diagram, which at low energies is equivalent to Weinberg-Tomozawa term, 
has a better ultraviolet behaviour.  

\begin{figure}[t]
\epsfig{file=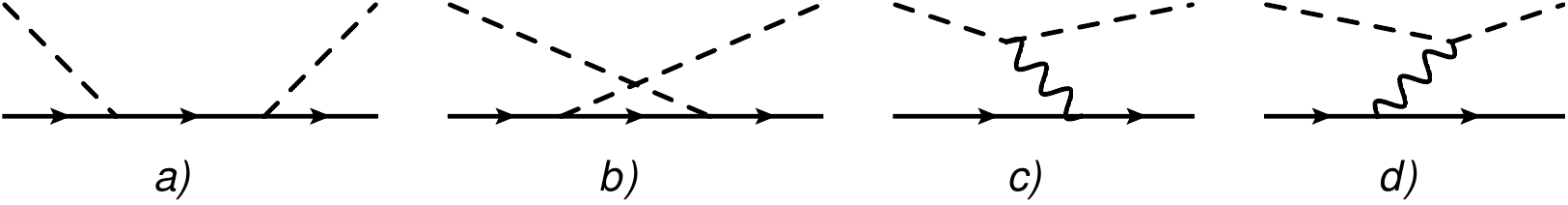,width=0.7\textwidth}
\caption[]{\label{LOPot} Time-ordered diagrams contributing to
  the LO meson-baryon potential. 
The solid, wiggled and dashed lines correspond to baryons, vector mesons and pseudoscalar mesons, respectively.}
\end{figure}

The LO $\pi N$ potential is given by time-ordered diagrams shown in Fig.~\ref{LOPot}.  Notice that while
we include the vector mesons as explicit degrees of freedom, for  pion-nucleon scattering for
small Mandelstam $t$ all four components of the momenta $q^\mu$ carried by vector meson lines are small
compared to their masses\footnote{This also applies to loop momenta, because after renormalization they
  are effectively cut off at small scales.}. Therefore, in the propagator of the vector meson
$\sim g^{\mu\nu}-q^\mu q^\nu/M_\rho^2$, the contribution of the second term is suppressed compared to the first
one. Thus, we include only the first term in the leading order 
potential by treating the second term as a higher order correction. This issue is discussed in more
detail in App.~\ref{app:A}. 
In TOPT, this leads to the standard rules (i.e.~similar to the
ones for scalar particles) for intermediate states containing vector meson lines.
Let us emphasize that this is completely different from  processes involving external vector mesons,
where such an approximation is not justified~\cite{Gulmez:2016scm,Du:2018gyn}.

By taking into account the projectors $P_+$ which reduce the expressions corresponding to the diagrams
of Fig.~\ref{LOPot} to the LO contributions to the effective potential, we have
(to obtain the amplitude/potential, one factor of $i$ is dropped in the expressions of the diagrams): 
\begin{equation}
V^{ba}_{\pi N} = 
\delta^{ba} \left[ g_V^+
+ i \,\vec \sigma \cdot(\vec q_2\times \vec q_1) 
 h_V^+  \right] + i\,\epsilon^{bac} \tau^c \left[ g_V^-
+ i \,\vec \sigma \cdot(\vec q_2\times \vec q_1) 
 h_V^-  \right] ,
 \end{equation}
 with 
 \begin{equation}
 	g_V^{\pm} = g_{V_a}^{\pm}+g_{V_b}^{\pm}+g_{V_{c+d}}^{\pm}, \quad 
 	h_V^{\pm} = h_{V_a}^{\pm}+h_{V_b}^{\pm}+h_{V_{c+d}}^{\pm},
 \end{equation}
 \begin{eqnarray}
   \label{potential}
 g_{V_a}^{\pm} & = & \vec q_1\cdot \vec q_2 \, h_{V_a}^{\pm}\, =\,  \frac{ g_A^2 m_N}{4
   F_\pi^2} \frac{\vec q_1 \cdot\vec q_2 }{\omega
   \left(p_1 + q_1,m_N\right) \left(\omega \left(p_1 + q_1,m_N\right)- E -i\,\epsilon \right)},\nonumber\\
 	g_{V_b}^+ &= & - g_{V_b}^- \,=\, - \vec q_1 \cdot\vec q_2\,
                       h_{V_b}^+ = \vec q_1 \cdot\vec q_2\, h_{V_b}^- \nonumber\\
 	&=&   \frac{ g_A^2 m_N}{4
   F_\pi^2} \frac{\vec q_1\cdot \vec q_2}{\omega \left(p_1-q_2,m_N\right)
   \left(\omega \left(p_1-q_2,m_N\right)+\omega(q_1,M_\pi)+\omega(q_2,M_\pi)-E -i\,\epsilon \right)},\nonumber\\
 	g_{V_{c+d}}^+ &=& h_{V_{c+d}}^\pm = 0,\nonumber\\
 	g_{V_{c+d}}^- &=& \frac{M_{\rho}^2\left(\omega(q_1,M_\pi)+\omega(q_2,M_\pi)\right)}{8F_\pi^2\,\omega(q_1-q_2,M_\rho)} 
 	\left(\frac{1}{\omega \left(p_2,m_N\right)+\omega
   \left(q_1-q_2,M_{\rho }\right)+\omega \left(q_1,M_{\pi
   }\right)-E-i\,\epsilon } \right.\nonumber\\
   &&\left.+ \frac{1}{\omega \left(p_1,m_N\right)+\omega
   \left(q_1-q_2,M_{\rho }\right)+\omega \left(q_2,M_{\pi
   }\right)-E-i\,\epsilon }\right).
 \end{eqnarray}
The actual calculations described below are performed in the CMS with
$\vec p_1 = - \vec q_1 = \vec p$  and $\vec p_2 = - \vec q_2 = \vec
p\, '$.

\subsection{Renormalization}

We work in the partial wave basis and write the leading order
potential as the sum of  the one-nucleon reducible and irreducible
parts,   
\bea
\label{LOP}
V _0 = V_{R} + V_I ,
\eea
where $V_{R} =V_a$ and  $V_I=V_b + V_{c+d}$.
For the above potential it is possible to write the solution to the
LO equation in a form (analogously to Ref.~\cite{Kaplan:1996xu}), 
that allows one to carry out a subtractive renormalization.

\noindent
To that end, we write the solution to the LO equation  as
\cite{Epelbaum:2015sha}
\begin{equation}
T_0=T_I+(1+T_I\,G)\,T_R (1+G\,T_I).
\label{OEQ12}
\end{equation}
Here and in what follows, we use a symbolic notation and do not
explicitly write the momentum integrations.
The amplitudes $T_I$  and $T_R$ satisfy the equations
\begin{equation}
T_I=V_I+V_I\,G\,T_I\,
\label{OEQ1}
\end{equation}
and
\begin{equation}
T_R=V_R+V_R\,G\, (1 + T_I G) \,  T_R\,.
\label{OEQ2}
\end{equation}
Notice that while the amplitude $T_I$ is finite in the removed regulator limit, it gets large finite
contributions. 
For example, the one-loop diagram with the iterated rho-meson-exchange
potential contains pieces which violate the  chiral  power counting. 
Such large power-counting-breaking contributions can
(and must) be systematically removed by additional finite
subtractions. We choose the subtraction scheme such that our iterated
amplitude matches the perturbative one obtained in  
chiral EFT with the EOMS renormalization scheme \cite{Fuchs:2003qc}. To implement such subtractions,
we apply a subtractive renormalization scheme analogous to the one of
Ref.~\cite{Epelbaum:2020maf}, adjusted to the pion-nucleon system. In particular, working in the CMS,
 we replace the pion-nucleon propagator $G(E)$
 with the subtracted propagator
$G_S(E)=G(E)-G(m_N)$. 
As discussed in Ref.~\cite{Epelbaum:2020maf}, this corresponds to taking into account contributions
of an infinite number of pion-nucleon counterterms. Notice that such
extra subtractions have no influence on the dynamical generation of
resonances or bound states, see App.~\ref{app:B} for details. 
Thus, instead of Eq.~(\ref{OEQ12}), we have
\begin{equation}
T_0^S=T_I^S+\left(1+T_I^S\,G_S\right)\,T_R \left(1+G_S\,T_I^S\right),
\label{OEQ12R}
\end{equation}
where the subtracted amplitude $T_I^S$ satisfies the equation
\begin{equation}
T_I^S=V_I+V_I\,G_S\,T_I^S\,.
\label{OEQ1R}
\end{equation}
The reducible potential $V_R$ can, in the partial wave basis,  be written as 
\begin{equation}
V_R(E,p',p)= \xi^T(p')\, {\cal C}(E)\xi(p)~,
\label{nuCfact}
\end{equation}
where $\xi^T(q) : = (1, q)$ with $q \equiv | \vec q \,|$ and the 2$\times$2 matrix ${\cal
  C}(E)$ is obtained from the partial wave reduction of Eq.~(\ref{potential}). 
Then, the amplitude $T_R$ is also given in a separable form
\begin{equation}
T_R(E,p',p)= \xi^T(p') {\cal X}(E)\xi(p)\,,
\label{chiCfact}
\end{equation}
with
\begin{equation}
{\cal X}(E)= \left[{\cal C}^{-1}-\xi\,G_R\,\xi^T -  \xi\,G_R\,T_I^S G_R\, \xi^T \,\right]^{-1}.
\label{chisol}
\end{equation}
Thus the final expression for the amplitude $T$ has the form
\begin{equation}
T_0^S=T_I^S+(\xi^T+T_I^S\,G_R\,\xi^T)\,{\cal X} (\xi+ \xi \,G_R\,T_I^S)~.
\label{taup}
\end{equation}
In a close analogy to Ref.~\cite{Epelbaum:2015sha}, we apply subtractive renormalization, i.e.~all divergences in
all loop diagrams are subtracted and  the coupling constants are substituted by their renormalized, finite values.  
For the amplitude of Eq.~(\ref{taup}) this amounts to the procedure outlined below.
A straightforward ultraviolet power counting demonstrates that the amplitude $T_I^S$ as well as 
$\Xi^T (q') \equiv \xi^T +T_I^S\,G_S\,\xi^T$  and $\Xi(q) \equiv \xi
+\xi\, G_S\,T_I^S$ are finite while ${\cal X}(E)$ is divergent.
Renormalization is carried out by performing subtractions that
correspond to taking into account counterterms generated
by the renormalization of the nucleon mass and the pion-nucleon coupling
constant. That is, the dreesed nucleon propagator is enforced to have
a pole at the physical mass of the nucleon $m_N$, and the renormalized
pion-nucleon coupling is required to take its physical value  $g_A$.  
%

\begin{figure}[htb]
\epsfig{file=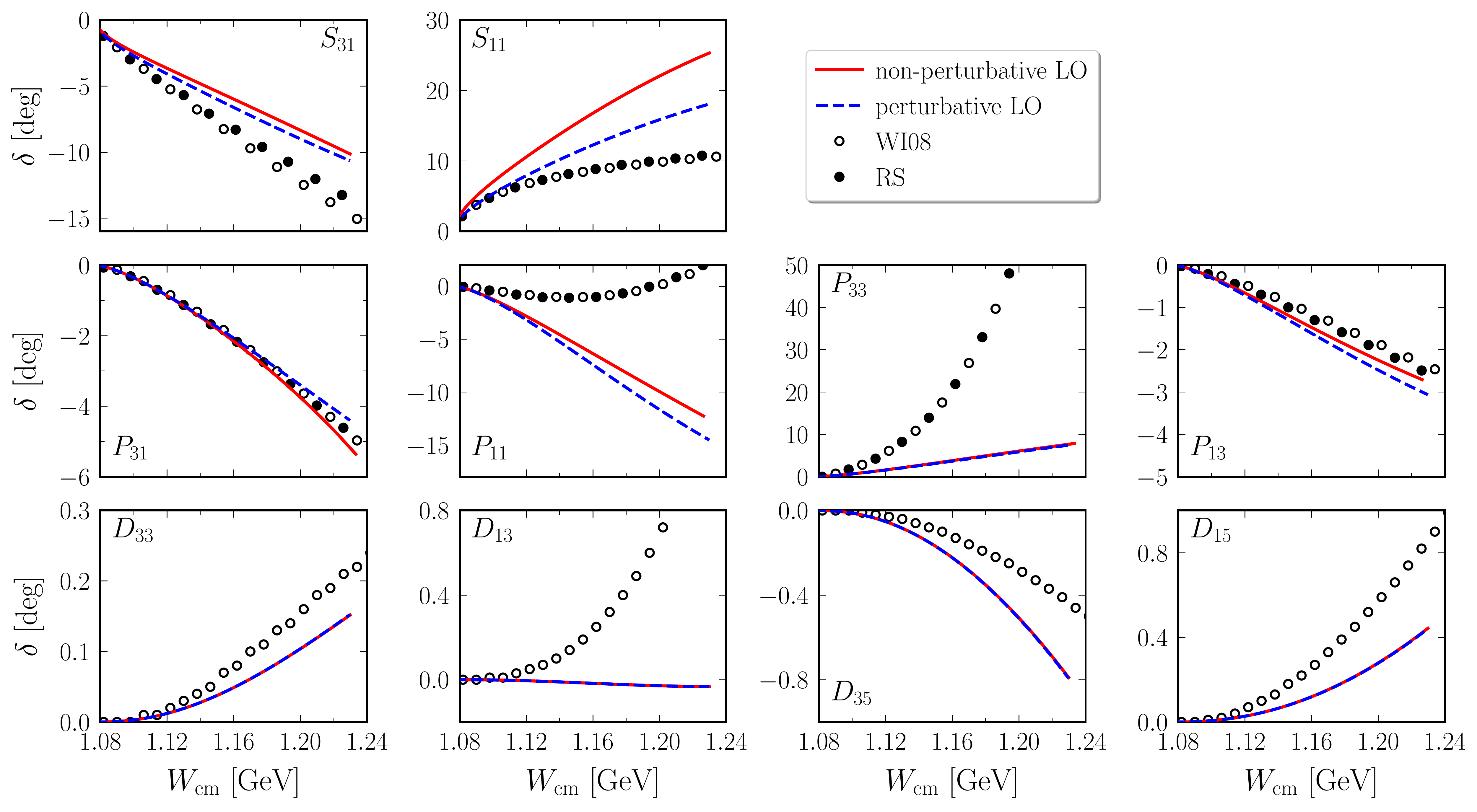,width=0.99\textwidth}
\caption[]{\label{LOPSh}  Pion-nucleon scattering phase shifts in standard partial-wave notation. 
  Blue (dashed) lines are the LO perturbative results  (tree-order result),
  red (solid) lines represent the results of the resummed LO potential,   
  while the dots and circles correspond to the Roy-Steiner analysis \cite{Hoferichter:2015hva}
  and GWU \cite{SAID}  phase shifts, respectively.}
\end{figure}

Our results for the pion-nucleon phase shifts based on the
renormalized amplitude are
shown in Fig.~\ref{LOPSh} in comparison with the ones obtained from a
perturbative tree-order calculation using the effective Lagrangiang
with vector mesons and the results from 
the Roy-Steiner equation analysis \cite{Hoferichter:2015hva} 
and the partial wave analysis of the George Washington University  group (GWU) \cite{SAID}. 
As expected for low-energies in the non-strange sector,
the results
for the renormalized resummed amplitudes are only
slightly different from the ones of the perturbative approach. Notice
that the $P_{33}$-wave can not be described
properly as long as the  $\Delta (1232)$ is not included as an
explicit degree of freedom, as it was already pointed out in
Sec.~\ref{effective_Lagrangian}.  An extension to the delta-full case
will be reported in a separate publication.

\section{Summary}
\label{conclusions}

In this paper we considered the meson-baryon scattering problem starting with
a manifestly Lorentz-invariant formulation of BChPT  and
applying time-ordered perturbation theory. 

We defined the effective potential
as a sum of  two-particle irreducible time ordered diagrams contributing to the meson-baryon scattering
amplitude. The full scattering amplitudes can be obtained by solving
the corresponding integral equations. By considering
an effective field theory of pseudoscalar and vector mesons and baryons,  
we obtained the integral equation for LO scattering amplitudes which is renormalizable.
By treating higher-order terms in the effective potential as perturbative corrections one can maintain
renormalizability also at higher orders.

The proposed approach for 
meson-baryon scattering in terms of the integral equations 
can provide quantitative information  on the convergence of ChPT in the single-baryon sector 
\cite{Mojzis:1999qw,Djukanovic:2004px,Mai:2009ce}.
ChPT provides a framework to perform perturbative calculation of the scattering amplitude order by order. 
It is, therefore, important to investigate the applicability of the chiral expansion, especially in the SU(3) sector,
where the perturbative expansion parameter is $m_K/\Lambda_\chi \sim 0.5$, with $m_K$ denoting the kaon mass
and $\Lambda_\chi$  the chiral symmetry breaking scale.  
In the chiral unitary approach proposed in this paper, iterations of the meson-baryon
scattering kernel within the integral equation
result in the nonperturbative resummation of a  certain class of renormalized contributions to the
scattering amplitude which are of higher orders according to the
chiral power counting.
Thus, by comparing the non-perturbative amplitude with its
perturbative expansion, one can get insights into the energy
region of the applicability of the chiral expansion. 
Similarly, since the scattering amplitude is a function of the light-quark masses, we can also
investigate the range of quark masses for which the chiral extrapolation of the lattice QCD data 
for meson-baryon scattering, see e.g.~Refs.~\cite{Torok:2009dg,Lang:2012db,Detmold:2015qwf,Lang:2016hnn,Andersen:2017una,Paul:2018yev}, can be trusted. 

As a first but still somewhat simplistic application we considered
here the pion-nucleon scattering amplitude and
compared the phase shifts obtained by solving the leading order integral equation to those of
chiral EFT. While the LO amplitude is finite in the removed cutoff
limit, it gets large  contributions that violate the
  chiral power counting in the low-energy region
and therefore requires finite subtractions. After
performing additional finite renormalization, the resummation
of an infinite number of higher order contributions is found to yield
small corrections to the phase shifts at low energies.
We note again that this approach is not applicable to all partial
waves since we have not included the $\Delta(1232)$
as an explicit degree of freedom. This can be done  
straightforwardly as it merely amounts to the corresponding extension
of the effective Lagrangian with no need to
modify the approach to calculate the scattering amplitudes described
in this work.

In view of the existing data on and the upcoming experiments of strangeness production,
applying our renormalizable framework to study the meson-baryon scattering in the SU(3) sector
will help to further understand the dynamics of hadrons with strangeness. In particular,
antikaon-proton scattering plays an important role in the study of the two-pole nature of the
$\Lambda(1405)$~\cite{Oller:2000fj,Jido:2003cb,Cieply:2016jby} and the properties of dense
nuclear matter, see~\cite{Tolos:2020aln} for a recent review.
Along this lines, we will carry out the leading- and next-to-leading order studies of the
meson-baryon scattering amplitudes in the strangeness $S=-1$ sector.

\section*{Acknowledgments}

This work was supported in part by BMBF (Grant No. 05P18PCFP1),  
by DFG and NSFC through funds provided to the
Sino-German CRC 110 ``Symmetries and the Emergence of Structure in QCD" (NSFC
Grant No.~11621131001, DFG Grant No.~TRR110), by the 
Georgian Shota Rustaveli National
Science Foundation (Grant No. FR17-354), by VolkswagenStiftung (Grant no. 93562)
 and by the CAS President's International
Fellowship Initiative (PIFI) (Grant No.~2018DM0034).

\appendix
\section{More on the iterated  $\rho$-meson propagator}
\label{app:A}

Here, we discuss in more detail the contribution from the longitudinal part
of the $\rho$-meson propagator. If we write the propagator of the $\rho$-meson as 
\begin{equation}
-i \,\frac{g^{\mu\nu}-\xi \, \frac{p^\mu p^\nu}{M_\rho^2}}{p^2-M_\rho^2}\,,
\end{equation}
 then the $\xi$-dependent part of the iteration of the $\rho$-exchange
diagram, i.e.~one-loop Lorentz-invariant box diagram, has the form
\begin{equation}
\frac{g^4 \xi  (\gamma \cdot q_1 +\gamma \cdot q_2)\left(\left(3 t \xi
  -12 M_\rho^2 (\xi -6)\right) B_0\left(t,{M_\rho}^2,{M_\rho}^2\right)-6 \xi  A_0\left({M_\rho}^2\right)
  +2 \xi  \left(t-6 {M_\rho}^2\right)\right)}{18 {M_\rho}^4},
\end{equation}
where the loop integrals are defined as
\bea
A_0(m^2)&=&\frac{(2\pi\mu)^{4-n}}{i\pi^2}\int\frac{{{\rm d}^nk}}{k^2-m^2}\ ,\nonumber\\
B_0(q^2,M_\rho^2,M_\rho^2)(s)&=& \frac{(2\pi\mu)^{4-n}}{i\pi^2}\int\frac{{{\rm d}^nk}}{[k^2-M_\rho^2][(k+q)^2-M_\rho^2]} .
\eea
As it is clearly seen from the above expression,  in the one-loop contribution to the scattering amplitude,
generated by the $\xi$-dependent terms, the pion and nucleon
denominators are cancelled and the obtained result
is polynomial in $t$ for $t \ll M_\rho^2$ and therefore can
be included in the renormalization of the contact interactions. Consequently,
the $\xi$-dependent term can be included in the higher order terms even when we iterate the $\rho$-meson
exchange diagram.

\section{On the generation of bound states or resonances}
\label{app:B}

Here, we want to discuss the issue of resonance generation in the presence of possible subtractions
in the integral equation.
Let us consider  a simple example. Suppose the potential is just a constant $C$, then the amplitude
depends only on the energy and the integral equation can be written as 
\begin{equation}
T(E)=C +  \int d^3 k \, C \, G(E,k) T(E)~.
\label{dem1}
\end{equation}
The solution of this equation is 
\begin{equation}
T(E) = \frac{1}{1/C -  \int d^3 k  \, G(E,k) }~.
\label{dem2}
\end{equation}
A resonance or bound state can be found by solving the equation
\begin{equation}
1/C - \int d^3 k  \, G(E,k)=0~.
\end{equation}
The integral which appears here is, however, divergent. We remove the divergence by renormalizing $C$ 
\begin{eqnarray}
T(E) & = & \frac{1}{\left[1/C -\int d^3 k  \, G(E_\mu,k)\right] - \int d^3 k  \, \left[ G(E,k) - G(E_\mu,k) \right]}
\nonumber\\
 & \equiv & \frac{1}{1/C_R(\mu)  -  \int d^3 k  \, \left[ G(E,k) - G(E_\mu,k) \right] }~,
\label{dem3}
\end{eqnarray}
where we subtract at $E= E_\mu$, with $E_\mu$ the renormalization scale, a convenient choice of which would
be e.g.~$E_\mu=E_{\rm threshold}-\mu$ for some non-negative $\mu$ of the order of the small scale in the probelm.
Obviously, such an identical transformation does not change the
position of the pole of the amplitude. i.e.~the resonance or bound
state. 
 
On the other hand, it is exactly equivalent to solving a subtracted integral equation,
\begin{equation}
T(E)=C_R+  \int d^3 k \, C_R \, \left[ G(E,k)-  G(E_\mu,k)\right] T(E)~,
\label{dem1R}
\end{equation}
which also takes into account contributions of an infinite number of counterterms, which are responsible
for the subtractions in each iteration of the equation.

\end{document}